\documentclass[conference]{IEEEtran}
\IEEEoverridecommandlockouts
\usepackage{cite}
\usepackage{amsmath,amssymb,amsfonts}
\usepackage{algorithmic}
\usepackage{graphicx}
\usepackage{textcomp}
\usepackage{xcolor}
\usepackage{float}

\usepackage{comment}

\def\BibTeX{{\rm B\kern-.05em{\sc i\kern-.025em b}\kern-.08em
    T\kern-.1667em\lower.7ex\hbox{E}\kern-.125emX}}
\begin{document}

\title{CAIA in Practice: Field Evaluation of an AI-Assisted Support System for Text-Based Online Counselling
}

\author{%
  \IEEEauthorblockN{Philipp Steigerwald\IEEEauthorrefmark{1},
    Nico Bienlein\IEEEauthorrefmark{1},
    Jennifer Burghardt\IEEEauthorrefmark{2},
    Mara Stieler\IEEEauthorrefmark{2},
    Robert Lehmann\IEEEauthorrefmark{2},
    Jens Albrecht\IEEEauthorrefmark{1}}%
  \IEEEauthorblockA{\IEEEauthorrefmark{1}Faculty of Computer Science,\\
    \IEEEauthorrefmark{2}Faculty of Social Sciences, Institute for E-Counselling\\
    Technische Hochschule Nürnberg Georg Simon Ohm, Nürnberg, Germany\\
    \{philipp.steigerwald,nico.bienlein,jens.albrecht,jennifer.burghardt,mara.stieler,robert.lehmann\}@th-nuernberg.de}%
}

\maketitle

\begin{abstract}
Rising global demand for mental health support creates significant service delivery challenges, with asynchronous email counselling serving as a crucial low-threshold channel for accessing care. This paper presents CAIA, a co-designed AI-based tool suite that demonstrates responsible AI integration into counselling practice through seven LLM-driven functions enhanced by retrieval-augmented generation. A field evaluation involved 34 professional counsellors conducting authentic sessions with trained student counsellees (36 threads, 321 messages, 1,257 AI outputs). User behaviour analysis confirms substantial adoption, revealing that professional autonomy and information accuracy are decisive for sustained acceptance, with counsellors particularly valuing interpretive functionalities that provide new perspectives and stimulate professional reflection.
\end{abstract}

\begin{IEEEkeywords}
AI-Assisted Counselling,
Email Counselling,
Large Language Models,
Retrieval-Augmented Generation,
User Adoption Study,
Human-AI Collaboration
\end{IEEEkeywords}

\section{Introduction}\label{sec:introduction}
Global mental health services struggle to meet escalating demand, while digital interventions emerge as promising solutions to bridge this gap. 
Text-based online counselling platforms offer accessible psychosocial support that can reach people who might not otherwise seek care \cite{torous_mental_2018,fulmer_artificial_2019,laranjo_conversational_2018}. These services enable help-seekers to interact with counsellors via text at any time and location, lowering barriers related to stigma \cite{borghouts_barriers_2021,brown_ai_2021,siddals_it_2024}, cost \cite{lattie_overview_2022,gamble_artificial_2020} and geographic distance \cite{negaro_geographic_2023,li_systematic_2023}. 
Asynchronous email counselling, in particular, has become a pivotal low-threshold channel for counsellees who prefer text-based, time-shifted interaction \cite{world_health_organization_mental_2023,bradley_e-mail_2011}.

Large language models (LLMs) present potential to enhance counselling services by automatically extracting, condensing and processing information from counselling communications \cite{adhikary_exploring_2024, so_aligning_2024, ohse_gpt-4_2024,stieler_technologische_2025}. 
AI assistance could enable counsellors to save time on information processing while simultaneously allowing them to engage more deeply with individual cases. 
By off-loading most of this analytical work, AI tools can free counsellors to invest more attention in the counselling relationship, which may in turn improve outcomes in online mental health services.
Given the inevitable integration of AI in mental health practice, the question is no longer \textit{whether} to employ such tools, but \textit{how} to integrate them responsibly while mitigating risks in this vulnerable domain.

\begin{figure}[t]
    \centering
    \includegraphics[width=1\linewidth]{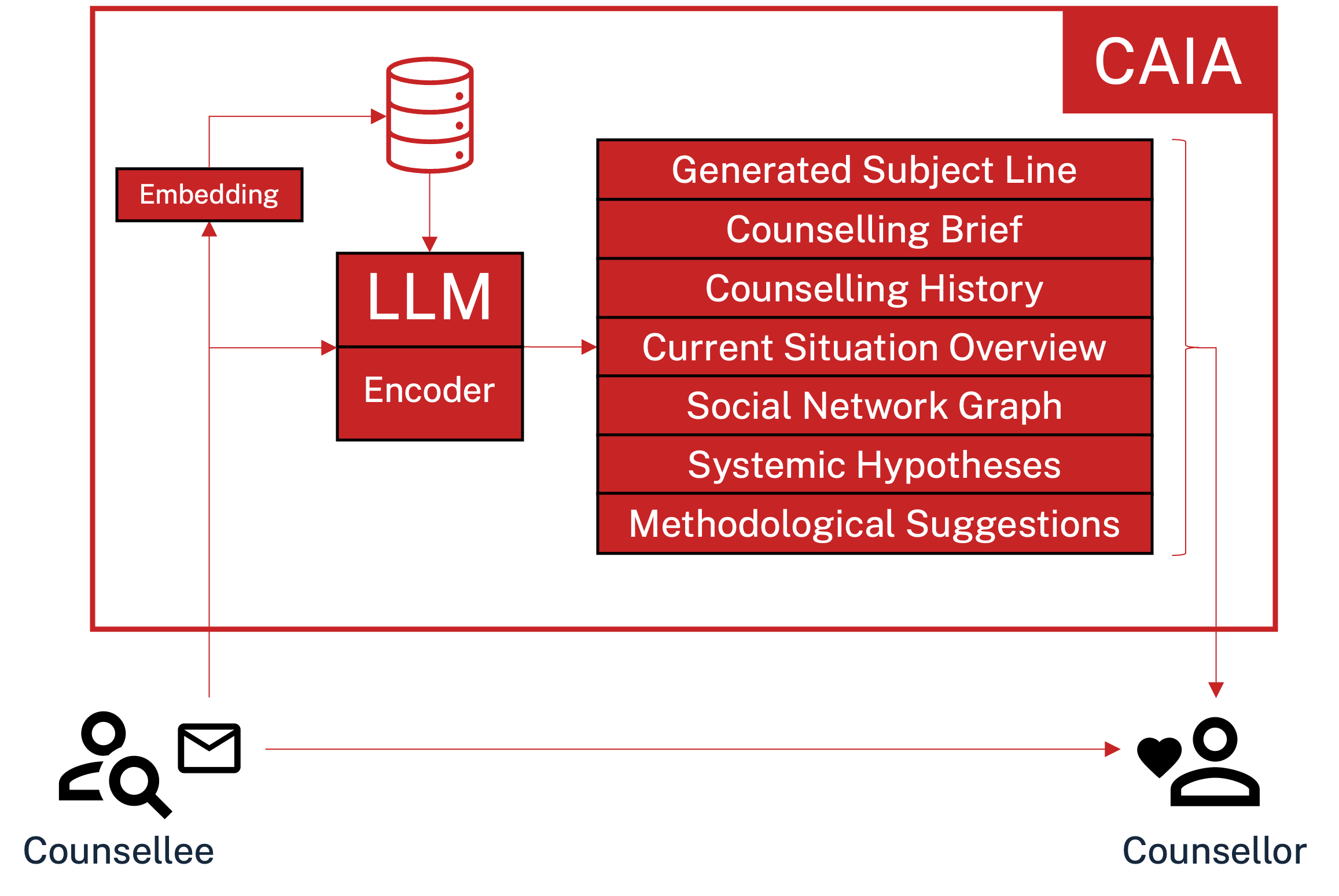}
    \caption{CAIA System}
    \label{fig:CAIA}
\end{figure}

Despite extensive development activities and various approaches to integrating LLMs into email-based counselling, limited knowledge exists about whether and how functions such as summarisation, information processing and visualisation are actually adopted by counsellors in practice \cite{guo_large_2024}. 
Understanding this intersection between AI capabilities and counselling practice requires developers to go beyond technical specifications and consider broader societal contexts, which is crucial for responsible technology development in mental health contexts \cite{murray-rust_ethicamanuensis_2022}.

This paper presents \textbf{CAIA} (Counselling AI Assistant), an AI-based tool suite designed to augment counsellors' skills without interfering with the traditional counselling relationship.
CAIA operates alongside the traditional email correspondence between counsellor and counsellee, providing AI-driven analytical functionalities as supplementary tools while preserving the integrity of the counselling exchange (as shown in Fig. \ref{fig:CAIA}). 
The system was developed through co-design involving professional counsellors, social science researchers and software developers to ensure alignment with professional practice requirements. 
CAIA can be deployed either as a fully fledged stand-alone counselling platform or as an API service that integrates seamlessly into existing systems.
To examine real-world adoption patterns, CAIA was deployed in a four-week field evaluation with 34 practising counsellors from three distinct cohorts: addiction counsellors, family and youth counsellors and newly qualified online counsellors. 
During this deployment, counsellors engaged with 13 trained students who enacted realistic case vignettes, resulting in 36 counselling threads and 321 messages between counsellors and counsellees.
The AI functionalities updated automatically at appropriate points as conversations progressed, generating 1,257 updates throughout the study period.

This work makes two key contributions to the intersection of AI and email counselling practice.

\begin{itemize}
    \item First, CAIA demonstrates how AI can be responsibly integrated into email counselling practice through co-designed, non-intrusive functionalities that preserve counsellor autonomy.
    \item Second, a comprehensive field evaluation that analyses counsellor adoption patterns and feedback, revealing which AI functionalities achieve professional acceptance and identifying key factors for successful integration into counselling workflows.
\end{itemize}

\section{Related Work}\label{sec:related-work}
A growing body of scholarship explores how artificial intelligence (AI) can \emph{support}—rather than replace—human counsellors during email and chat interventions. \textit{HAILEY} provides empathic rewrites for peer-support replies, achieving a 19.6\% increase in conversational empathy \cite{sharma_humanai_2023}. \textit{PARTNER} employs deep reinforcement learning to enhance empathy in posts \cite{sharma_towards_2021}, whilst \textit{CARE} couples large language models with Motivational Interviewing frameworks for context-aware guidance \cite{hsu_helping_2025}. Training tools such as \textit{ReadMI} provide real-time metrics to reduce cognitive load during skills practice \cite{hershberger_motivational_2024}. Safety-critical triage models automatically flag crisis messages, significantly reducing triage latency \cite{swaminathan_natural_2023}. These systems illustrate the breadth of AI-mediated functionality—empathic rewriting, strategic suggestion, skills feedback and risk triage—now available to augment counsellors in text-based practice.

However, field evidence on user adoption patterns remains limited. While \textit{HAILEY} demonstrated successful deployment in synchronous peer chat \cite{sharma_humanai_2023}, \textit{CARE} and \textit{PARTNER} have been evaluated chiefly in controlled settings \cite{hsu_helping_2025,sharma_towards_2021}. \textit{ReadMI} has shown promise educationally, yet routine adoption remains unexplored \cite{hershberger_motivational_2024}, whilst crisis-triage integration into counsellor workflows lacks systematic examination \cite{swaminathan_natural_2023}. Particularly for asynchronous email counselling, questions remain about professional counsellors' adoption of multifunctional AI assistance tools. This study addresses these gaps by presenting a field evaluation of CAIA, examining how professional counsellors engage with AI-powered functionalities during authentic counselling interactions in German-language email counselling practice.

\section{System Overview of CAIA}
\label{sec:system}
CAIA was developed through a comprehensive co-design process involving software developers, social science researchers and practising counsellors from the email counselling domain.
The system architecture, AI-driven functionalities and conventional tools underwent multiple development iterations informed by usability testing sessions and stakeholder workshops. 
The resulting system reflects domain expertise integration and validation through practitioner feedback across successive prototype refinements.

\subsection{Functionality Set}
Drawing from established needs in asynchronous text-based counselling practice, CAIA implements seven AI core functionalities (see Fig. \ref{fig:CAIA}):
\begin{itemize}
  \item \textbf{Generated Subject Line} – generates descriptive subject lines to supplement generic counsellee headings like ``Help'' or ``Problem'', presented alongside the original to preserve counsellee framing whilst improving navigation.
  
  \item \textbf{Counselling Brief} – extracts explicit and implicit help-seeking goals from counsellee messages. The system generates bullet points capturing what the counsellee seeks from the counselling relationship, with each goal supported by direct textual references.
  
  \item \textbf{Counselling History} – generates message-by-message summaries that capture the essential content of each communication in the counselling thread. The system creates bullet points that distil the core information. Each summary entry includes the message date, author identification and direct textual references.
  
  \item \textbf{Current Situation Overview} – extracts factual information about the counsellee's current life circumstances from their most recent communication. The system identifies situational facts including housing, educational, professional, family and health considerations, with each element supported by direct textual references.
        
  \item \textbf{Social Network Graph} – extracts and visualises the counsellee's social network, identifying entities, relationships and attributes like age or personality associated with each actor as described by the counsellee.
  
  \item \textbf{Hypotheses} – generates knowledge-based hypotheses about interpersonal dynamics and systemic patterns. Drawing from systemic counselling principles, the system formulates hypotheses in conditional language incorporating systemic analysis approaches.
        
  \item \textbf{Methodological Suggestions} – deliver context-aware guidance suggestions for shaping the next counselling step. Drawing on best-practice rules from online-counselling research, educational counselling and systemic counselling, the functionality offers phase-specific recommendations such as systemic questions, resource-oriented tasks and psycho-educational elements.

\end{itemize}

The first five functionalities primarily extract and summarise information from counsellee communications (information extraction functionalities), whilst the latter two interpret contextual patterns to generate professional insights about case dynamics and intervention strategies (interpretive functionalities).

These functionalities operate through an event-driven architecture with differential triggering patterns. The Generated Subject Line is computed once upon receipt of the counsellee's initial message and remains static throughout the counselling thread. The Counselling Brief, Current Situation Overview, Social Network Graph, Hypotheses and Methodological Suggestions are regenerated following each counsellee message, whilst the Counselling History updates after every message from both counsellee and counsellor to maintain comprehensive chronological documentation.

To ensure domain-aligned outputs, six of the seven functionalities are enhanced through a RAG pipeline providing domain-specific counselling knowledge. The Social Network Graph operates without knowledge augmentation, as initial evaluation revealed that external literature introduced spurious actors into the network visualisation.

In addition to the AI functionalities, CAIA provides two conventional tools in stand-alone mode:
\begin{itemize}
\item \textbf{Annotations} – enables counsellors to highlight and colour-code specific text passages within counsellee messages. Highlighted passages are consolidated in a dedicated panel, allowing easy review and one-click navigation back to the original text.
\item \textbf{Notes} – provides a conventional note-taking interface that enables counsellors to record observations, counselling insights and session planning notes. This functionality functions as a digital notepad integrated within the CAIA interface. During testing phases, it also serves as a channel for capturing feedback about AI functionalities.
\end{itemize}

\subsection{Knowledge Integration via RAG}
The Retrieval-Augmented Generation pipeline integrates domain-specific counselling knowledge through a two-tier document architecture. The knowledge base comprises a general superset containing universally applicable counselling principles and functionality-specific subsets providing targeted theoretical approaches. This separation addresses conflicting perspectives in counselling literature, allowing each functionality to draw from theoretically aligned sources whilst avoiding contradictory recommendations.

During functionality computation, the system retrieves relevant passages from both the global corpus and functionality-specific subset, combining established counselling principles with case-relevant details.

The knowledge base supports dynamic expansion and modification to accommodate evolving counselling approaches and emerging theoretical frameworks.

\begin{figure*}[htbp]
    \centering
    \includegraphics[width=1\linewidth]{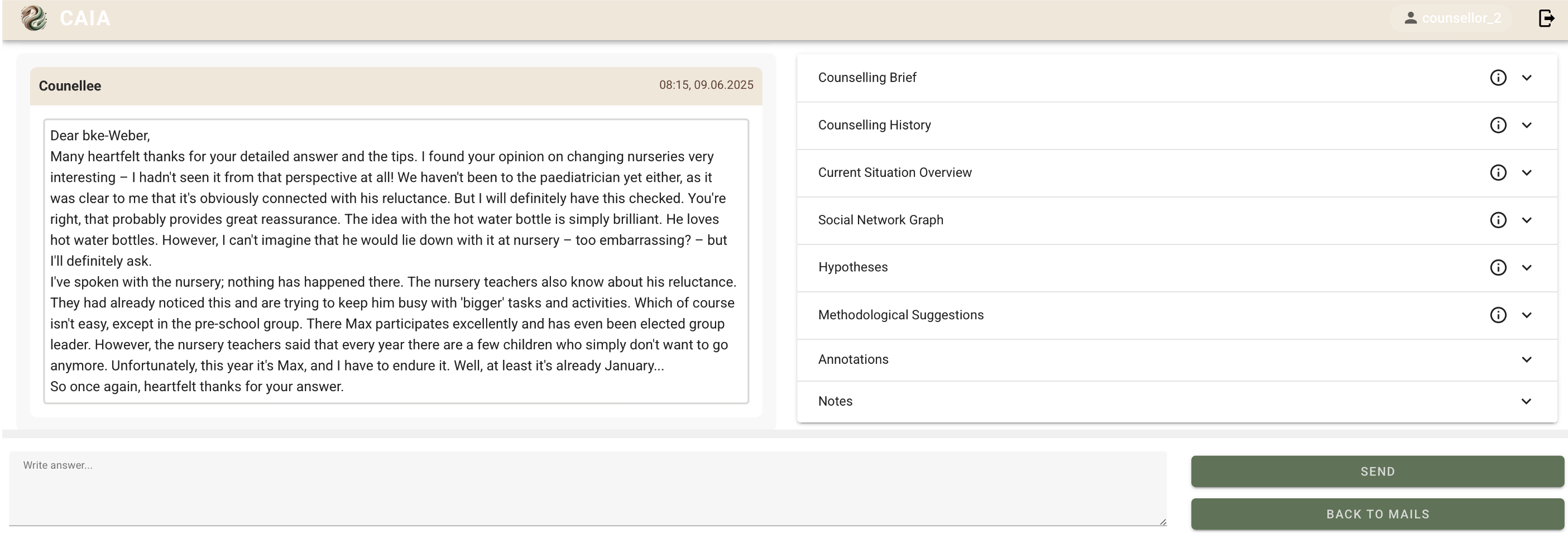}
    \caption{CAIA user interface displaying the split-screen design with email conversation thread (left) and expandable AI functionality panels (right).}
    \label{fig:ui}
\end{figure*}

\subsection{Architecture}
CAIA employs a distributed containerised architecture organised into three layers—a web frontend, a REST API for authentication and message handling and an asynchronous computation tier for AI processing. Deployment is orchestrated with Docker Compose, while Flask exposes the REST endpoints, Redis queues background LLM jobs and Celery executes those jobs asynchronously. Dialogue data and user metadata are stored in PostgreSQL, semantic embeddings for RAG retrieval reside in ChromaDB and an Nginx reverse proxy routes external traffic to the appropriate containers. This modular stack lets each component scale or swap out independently.

\textbf{Deployment Modes: }CAIA can operate in two distinct modes. 
In stand-alone mode it serves as a complete counselling platform, handling functions such as automatic client-to-counsellor allocation and load balancing. 
Alternatively, it can be deployed in API-service mode, allowing external email counselling systems to mirror email threads to CAIA via a REST endpoint, where CAIA performs the AI computations for each functionality and returns the processed results to the requesting system.

\textbf{Multi-Tenant Architecture: }The system supports multiple counselling institutes on a single deployment, each with customised configurations for different specialisations (e.g. addiction counselling, family counselling, debt counselling). Institute Administrators configure active functionalities, LLM endpoints, prompts and knowledge bases, while System Administrators provision new institutes via API.

\textbf{Asynchronous Processing:} The computation layer ensures ordered AI functionality generation through message-level sequential processing within email threads, while enabling concurrent functionality generation across multiple threads. This architecture supports diverse LLM deployments from cloud services to self-hosted inference servers with granular model assignment per functionality and institute.

\subsection{Optional User Interface}
The CAIA interface was designed through co-design with practitioners to ensure AI functionalities support rather than disrupt counselling workflows. 
The split-screen design reflects counsellors' preference for maintaining primary focus on the client conversation while accessing AI assistance only when needed. 
The email conversation thread occupies the left side, whilst functionalities are accessible through collapsible panels on the right (see Fig. \ref{fig:ui}). 

This non-intrusive approach emerged from co-design feedback emphasising that AI tools should never impose themselves on the counselling process. 
Counsellors can choose which functionalities to engage with and when, preserving professional autonomy. Only one panel expands at a time to maintain focus.

\begin{figure}[htbp]
    \centering
    \includegraphics[width=1\linewidth]{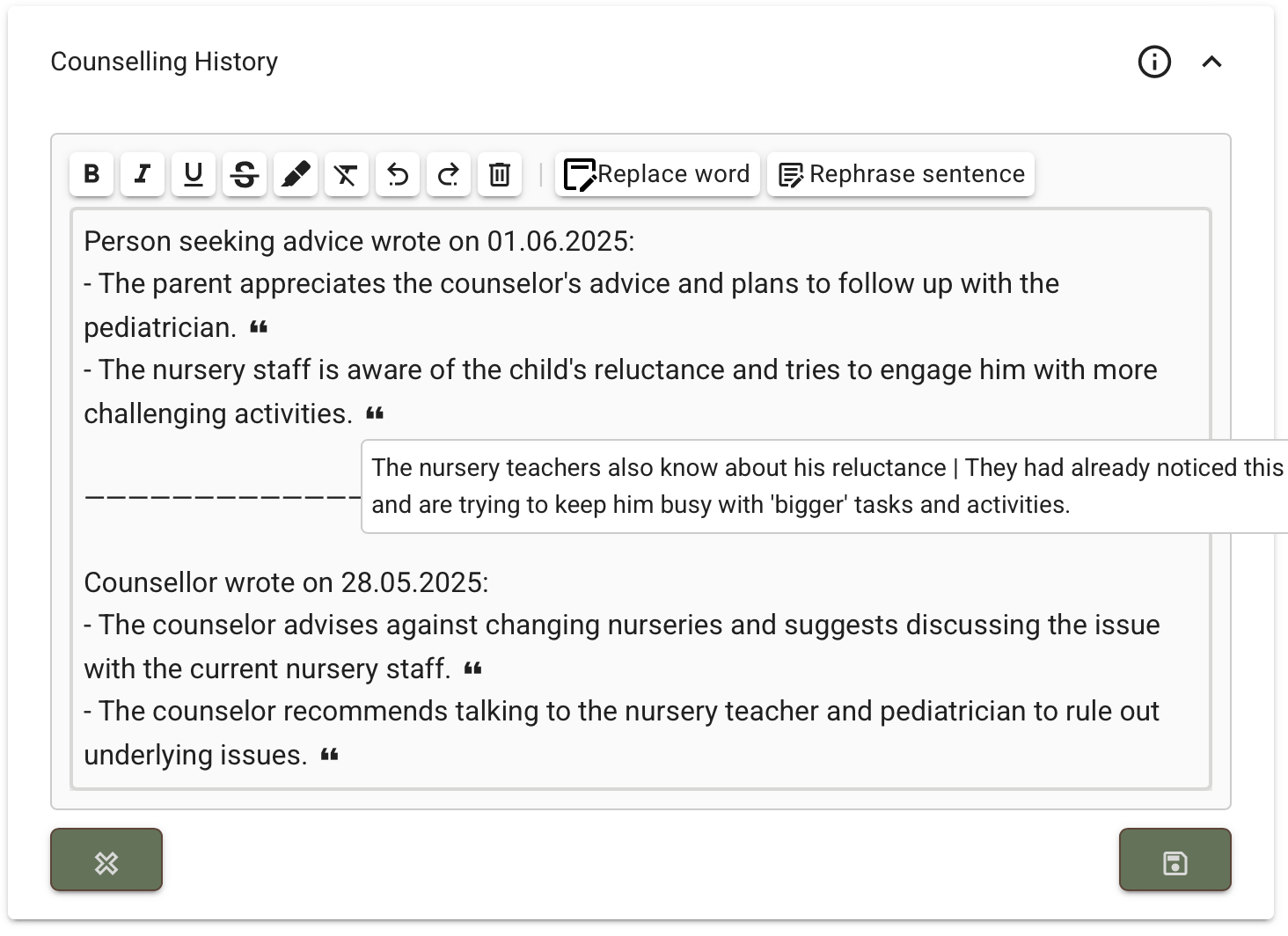}
    \caption{Counselling History functionality panel in edit mode, displaying the editing toolbar with formatting options, AI-powered text manipulation functions (replace word, rephrase sentence) and reference symbols with hover tooltips showing source passages.}
    \label{fig:edit}
\end{figure}

Each functionality displays its generation history through visual separators (horizontal dashed lines) to distinguish between content versions as cases evolve (see Fig. \ref{fig:edit}). 
The newest content always appears at the top for immediate visibility, whilst previous versions remain accessible below for transparency and tracking analytical development over time.

All functionalities support editing capabilities through integrated toolbars, enabling counsellors to refine AI-generated content according to their professional judgment. 
The editing interface includes standard formatting tools and AI-powered functions for contextual word replacement and sentence reformulation. 
This collaborative human-AI approach ensures technological assistance enhances rather than replaces counsellor expertise, preserving counsellor autonomy.

For functionalities that provide textual references, the system displays bold quotation marks after relevant bullet points (visible in Fig. \ref{fig:edit}). 
Hovering over these reference symbols reveals the source passage, whilst clicking automatically navigates to the original text in the counsellee's message.

\section{Field Test}
\label{sec: field test}
CAIA was deployed in a comprehensive field evaluation to examine professional counsellors' adoption of AI assistance functionalities during counselling interactions, capturing usage patterns in professional practice settings.



\subsection{Participants}
The field evaluation involved 34 professional counsellors across two domains of online counselling. The sample comprised three groups: nine experienced counsellors from bke (Germany's national association for educational and family counselling) assigned family and youth counselling cases, 14 experienced counsellors from DigiSucht (a federally funded digital addiction counselling platform) assigned addiction counselling cases and 11 newly qualified counsellors who had recently completed a Certificate Course in Online Counselling at the Institute for E-Counselling at TH Nürnberg assigned family and youth counselling cases. For the evaluation, 13 trained students took on counsellee roles using realistic case vignettes, with each student corresponding with two or three counsellors simultaneously and freely interpreting the case vignettes to create the most realistic and authentic cases possible.

\subsection{Study Design}
The study employed a naturalistic field evaluation design conducted over four weeks, with CAIA accessible around the clock. Participating counsellors were introduced to the system through a walkthrough and video, then instructed to conduct counselling sessions as usual while exploring CAIA's AI tools and documenting feedback in the notes panel.

Each institute received domain-specific configurations. Bke counsellors and newly qualified counsellors used prompts and RAG documents optimised for family and youth counselling, while DigiSucht operated with addiction counselling-specific configurations. All AI generation used mistralai/Mistral-Small-3.1-24B-Instruct-2503 deployed on self-hosted German servers for data sovereignty.

\subsection{Data Acquisition}
The study utilised 36 authentic online counselling email threads, generating 321 messages (180 from counsellees, 141 from counsellors). Securing realistic counselling interactions while maintaining ethical standards represents a methodological achievement given the sensitive nature of such communication.

Comprehensive behavioural data was captured through integrated system logging and Matomo tracking (on-premise), recording all user interactions without disrupting the counselling process. Data collection encompassed how often each functionality panel was opened, how long it stayed visible, the paths counsellors took when navigating from one panel to another and which panels remained open while they composed replies.

Ethical considerations included informed consent, data anonymisation protocols and secure data handling consistent with EU data protection requirements. The study design prioritised participant autonomy, with counsellors free to use or ignore AI tools according to their professional judgment.

\section{User Behaviour Analysis}
Across the 36 counselling threads, CAIA produced 1,257 AI outputs during the evaluation period. The behavioural analysis covers 31 of the 34 participating counsellors (three declined browser-based tracking), logging 1,128 panel openings across the AI functionalities and two standard tools. Generated Subject Line usage was not traceable as it displays automatically.

\subsection{Frequency and Engagement Intensity}
Clear usage preferences emerged among CAIA's functionalities. 
Methodological Suggestions was most frequently accessed (190 openings, 16.8\%), followed by Hypotheses (178 openings, 15.8\%). 
The middle tier comprised Counselling History (15.0\%), Counselling Brief (13.1\%), Current Situation Overview (12.7\%), Social Network Graph (11.7\%) and Notes (11.7\%).
Annotations showed minimal engagement with only 36 panel openings (3.2\%).

\begin{figure}[htb]
    \centering
    \includegraphics[width=1\linewidth]{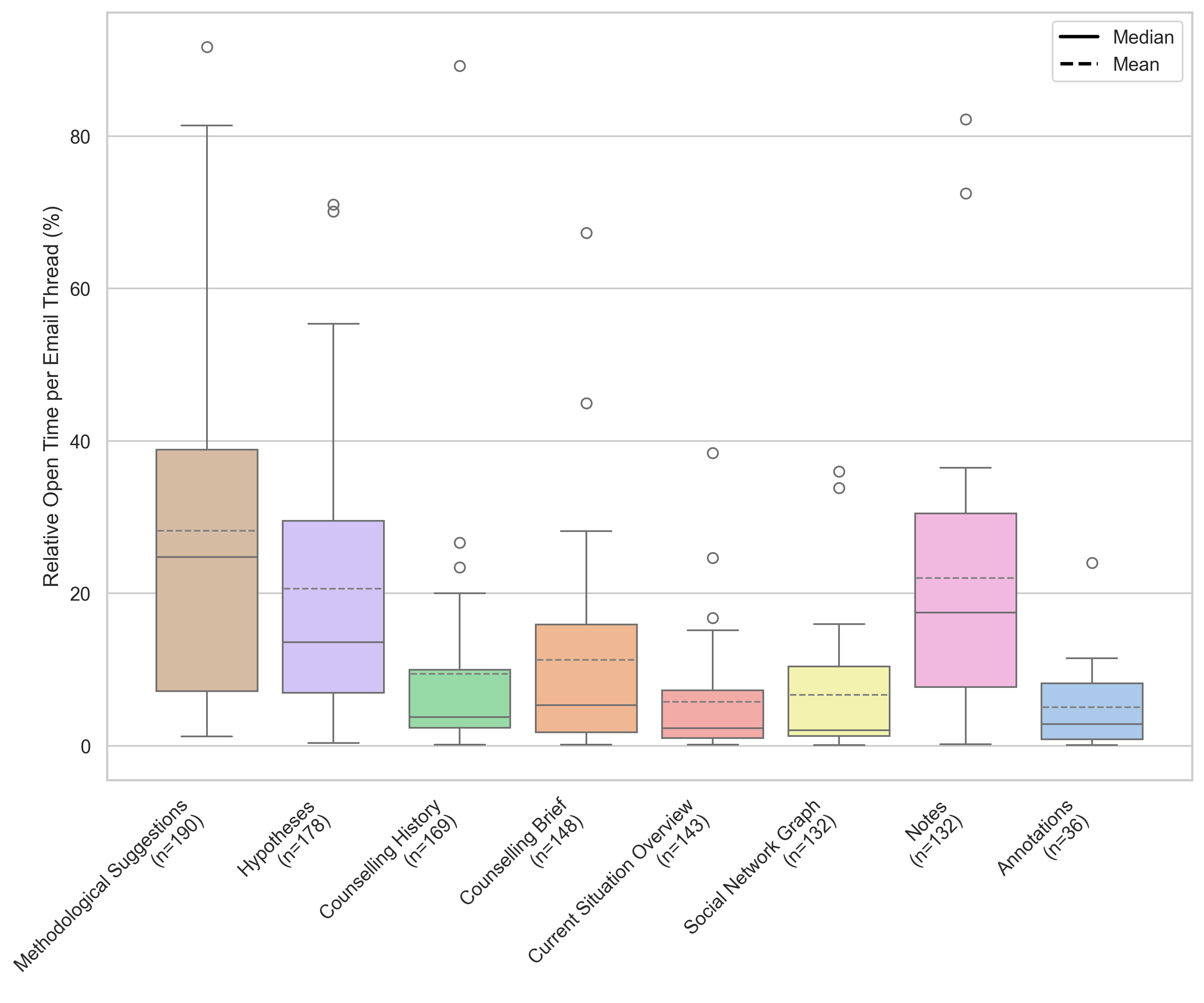}
    \caption{Boxplots of each functionality’s relative open-time per email thread (labels show total opens $n$).}
    \label{fig: relative_share_functionality_count}
\end{figure}

Figure \ref{fig: relative_share_functionality_count} shows, for each functionality, the distribution of its share of panel-open time within a mail thread (labels indicate total openings). Methodological Suggestions and Hypotheses top the ranking, each averaging more than 20\% of thread time.

\begin{figure*}[ht]
    \centering
    \includegraphics[width=1\linewidth]{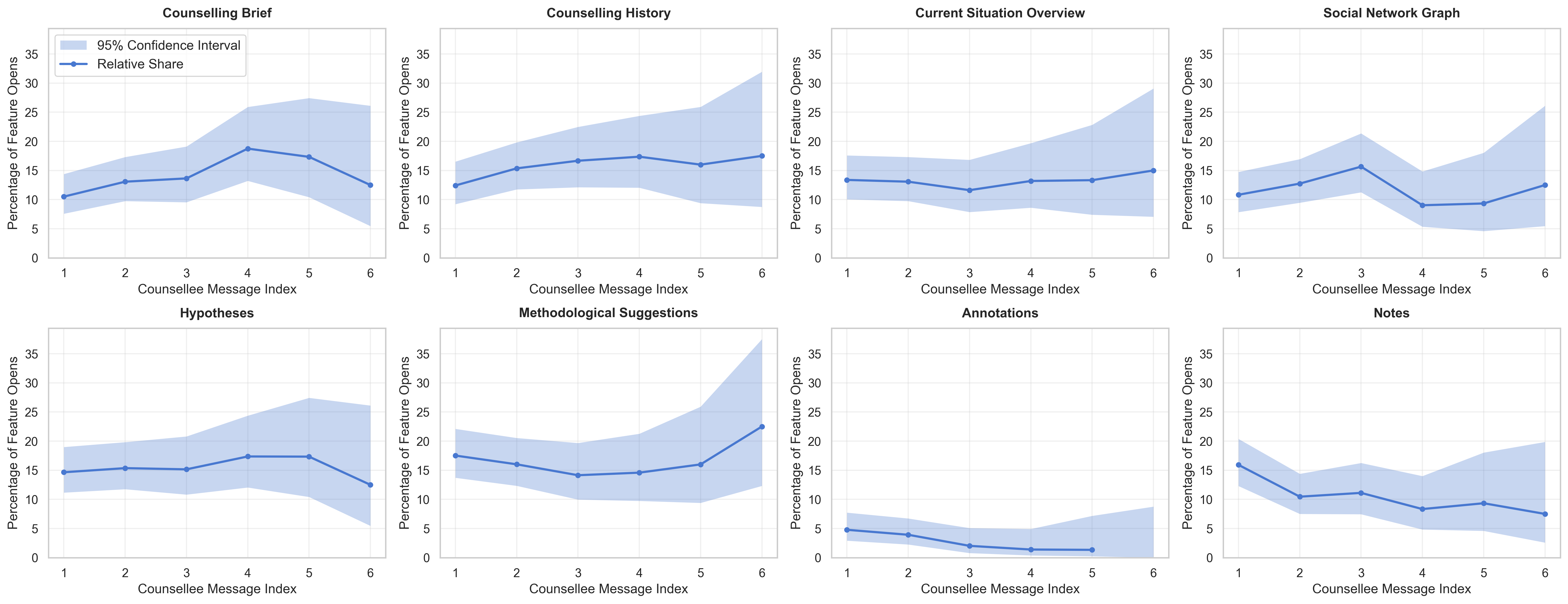}
    \caption{Functionality usage trends across counselling progression showing relative functionality engagement patterns with 95\% confidence intervals. Each subplot displays the percentage of functionality opens at different counsellee message indices (1-6), revealing how functionality preferences evolve throughout the counselling process.}
    \label{fig:functionality_trends}
\end{figure*}

Analysis of functionality panel openings across counselling progression reveals that while functionalities show similar initial usage frequencies, distinct utilisation patterns emerge as cases develop (Fig. \ref{fig:functionality_trends}). 
The Counselling Brief shows a curvilinear pattern—rising mid-stage before declining as help-seeking goals become clarified.
Counselling History demonstrates steady upward growth, providing updated summaries after each new message.
Current Situation Overview maintains stable utilisation around 14-15\%. 
The Social Network Graph shows initial engagement that subsequently declines.
Hypotheses usage rises modestly then settles around 13\% in longer exchanges, as initial hypotheses solidify over time.
Methodological Suggestions fluctuate with initial average usage, mid-stage decline, then increased utilisation in complex cases potentially requiring alternative intervention strategies.
Documentation functionalities show declining patterns, with Annotations decreasing from 5\% to 2\% and Notes from 16\% to 8\%.


\subsection{Functionality Transition Analysis}
\begin{figure}[H]
  \centering
  \includegraphics[width=\linewidth]{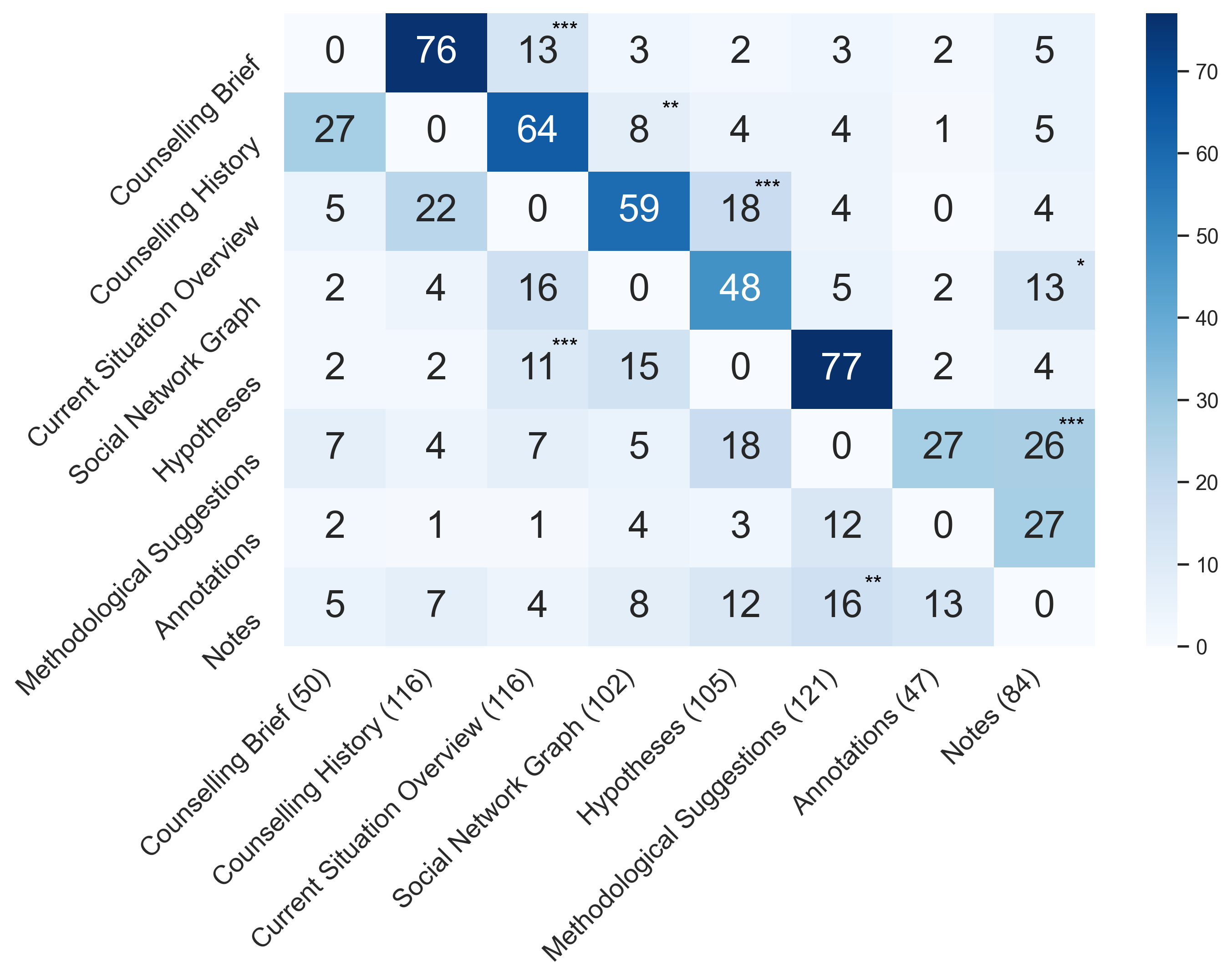}
  \caption{Functionality transition heat-map showing counsellor navigation between CAIA components. Counts are printed in each cell, with asterisks denoting significantly over-represented transitions (* $p_{\text{adj}}{<}.05$, ** $p_{\text{adj}}{<}.01$, *** $p_{\text{adj}}{<}.001$).}
  \label{fig:usage_heatmap_sig}
\end{figure}

Across the 741 recorded panel transitions (Fig. \ref{fig:usage_heatmap_sig}) counsellors largely opened the panels in the order presented in the interface. 
This orderly pattern appears as two slender diagonals in the heatmap - the band just above the main diagonal shows successive moves to the next panel, while the mirrored band below marks returns to previously opened panels. 
To detect purposeful shortcuts beyond sequential browsing, the three central diagonals were excluded and an independence test was applied to the remaining 42 cells. 
Expected counts were derived from the row and column marginals, standardised Pearson residuals were computed and two-sided $p$-values were adjusted with the Benjamini–Hochberg procedure. 

Seven transitions are significantly over-represented, comprising two bidirectional feedback loops. Current Situation Overview $\leftrightarrow$ Hypotheses (18***/11***) indicates iterative analytical engagement where counsellors cycle between factual evidence and interpretation, while Methodological Suggestions $\leftrightarrow$ Notes (26***/16**) reveals a distinctive feedback loop where counsellors move from reviewing AI suggestions to providing feedback in the notes panel, then return to the methodological guidance for further consideration, systematically bypassing the Annotations functionality in this iterative evaluation process.

Beyond these analytical cycles, three shortcuts reveal purposeful navigation patterns. First, Counselling Brief $\rightarrow$ Current Situation Overview (13***) skips Counselling History, suggesting counsellors cross-reference client requests against current circumstances for analytical validation. Second, Counselling History $\rightarrow$ Social Network Graph (8**) skips Current Situation Overview, likely reflecting verification of newly mentioned entities from the history against existing network representations. Third, counsellors jumped from Social Network Graph $\rightarrow$ Notes (13*) indicating immediate feedback documentation.

\subsection{Consultation During Response Composition}
Analysis of typing session dynamics examines which functionalities counsellors actively referenced whilst composing responses to counsellees. 
\begin{figure}[!ht]
    \centering
    \includegraphics[width=1\linewidth]{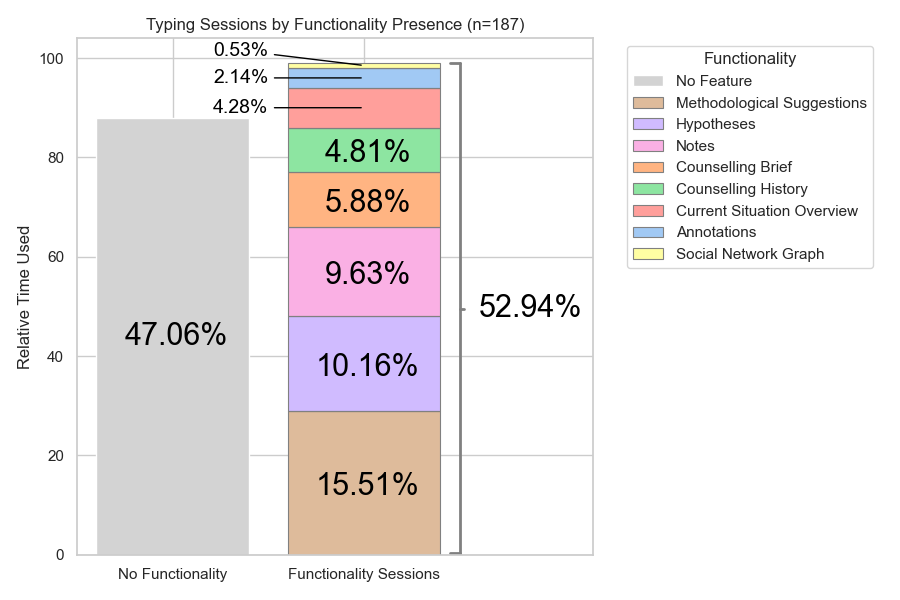}
    \caption{Distribution of functionality panel states during counsellor typing sessions showing active consultation patterns}
    \label{fig:relativ typing}
\end{figure}
In over half (52.94\%) of all typing sessions, counsellors maintained at least one functionality panel open while crafting replies (see Fig. \ref{fig:relativ typing}). 
The information extraction functionalities collectively accounted for only 17.64\% of typing sessions. 
Notes ranked higher with 9.63\%, potentially reflecting either active reference or simply leaving this panel open as the last interface element. 
Most notably, the interpretive functionalities collectively accounted for 25.67\% of typing sessions, suggesting counsellors actively engaged with these analytical insights during response composition.

\subsection{Insights from Counsellor Feedback}
\label{sec:qualitative_insights}

This analysis examines 58 counsellor feedback segments collected in the notes panel during the four-week evaluation, covering six of the seven AI functionalities listed in Table\ref{tab:feedback_distribution} (no feedback was received for the Generated Subject Line).
\begin{table}[htbp]
\centering
\caption{AI Functionality Feedback Distribution}
\label{tab:feedback_distribution}
\begin{tabular}{|l|c|c|}
\hline
\textbf{AI Functionality} & \textbf{Positive} & \textbf{Negative} \\
\hline
Counselling Brief & 5 & 4 \\
Counselling History & 5 & 1 \\
Current Situation Overview & 3 & 2 \\
Social Network Graph & 1 & 14 \\
Hypotheses & 3 & 7 \\
Methodological Suggestions & 8 & 5 \\
\hline
\textbf{Total} & \textbf{25} & \textbf{33} \\
\hline
\end{tabular}
\end{table}

Information extraction functionalities received positive feedback for supporting counselling practice, with counsellors appreciating they were \textit{``compact and helpful''} and found summaries \textit{``well put together.''}
However, negative feedback consistently focused on accuracy issues, with counsellors reporting the AI \textit{``entered information with no basis in the message content''} and found extractions \textit{``incomplete, overlooking crucial information,''} undermining professional confidence.

The Social Network Graph received the most negative feedback (14 negative vs. 1 positive), confirming declining usage over time.
While counsellors appreciated the visualisation concept for providing \textit{``a quick outside view of the system''} implementation errors severely undermined its utility.
Counsellors reported systematic problems where \textit{``roles were assigned incorrectly''} and noted \textit{``duplicate institutions where only one existed''} with these content inaccuracies alongside technical problems overshadowing the conceptual value.

Interpretive functionalities showed high behavioural engagement with mixed qualitative feedback.
Hypotheses presented a complex adoption pattern, with counsellor notes revealing both agreement and disagreement with AI-generated perspectives.
While some counsellors noted \textit{``the hypotheses are very interesting and also helpful,''} others found them \textit{``superficial''} or criticised them as \textit{``not professionally grounded.''}
Methodological Suggestions received predominantly positive responses for providing \textit{``useful confirmation and fresh impulses''} with counsellors reporting they could \textit{``easily just start writing''} and gained \textit{``ideas for additional suggestions in responses''}.
Although some counsellors mentioned inappropriate suggestions and contradictions with professional standards, overall feedback emphasised the functionality's value in inspiring new approaches and easing response composition.

\subsection{Overall Adoption Patterns}
Across behavioural and qualitative data, distinct patterns emerged for each functionality, revealing selective integration of AI tools based on professional value.

\textbf{Information extraction functionalities} demonstrated consistent adoption patterns across analytical dimensions.
These functionalities occupied the middle tier of usage frequency with brief engagement sessions, showed purposeful cross-referencing patterns and minimal utilisation during response composition.
Qualitative feedback revealed a clear dichotomy: counsellors appreciated their compact design and practical utility for rapid case orientation, but criticised accuracy issues including hallucinations and incomplete information extraction undermining professional confidence.
This suggests they function optimally as orientation tools rather than active consultation resources, with greatest potential value in case handovers or returning to cases after extended periods.

The \textbf{Social Network Graph} showed similar initial opening frequency but exhibited declining usage, frequent transitions to Notes indicating error documentation and minimal utilisation during response composition.
Qualitative feedback was negative, with counsellors appreciating the visualisation concept but reporting systematic technical problems.
This suggests that while the concept may hold value for systemic perspectives, the current implementation requires fundamental redesign—potentially through task decomposition, enhanced processing pipelines and robust language models—to achieve necessary accuracy and reliability.

\textbf{Hypotheses} presented a complex adoption pattern, ranking second in frequency with high engagement intensity, extended viewing times, bidirectional transitions with Current Situation Overview indicating hypothesis validation and frequent consultation during response composition.
Qualitative feedback revealed a paradox: while generating more negative than positive responses, counsellors simultaneously valued their capacity to stimulate reflection despite finding them superficial.
This contradiction—high engagement despite negative feedback—suggests hypotheses serve their intended purpose of promoting professional discourse, indicating that challenging AI content holds value even when critically evaluated.

\textbf{Methodological Suggestions} emerged as the clear leader across all analytical dimensions, showing highest opening frequency, longest session duration, bidirectional transitions with Notes suggesting feedback loops and most active consultation during response composition.
Qualitative feedback was predominantly positive, with counsellors valuing fresh impulses without imposed solutions, though some noted inappropriate suggestions and contradictions with professional standards.
This suggests successful integration into active counselling work, indicating that AI tools providing interpretive guidance may be embraced when they enhance rather than constrain autonomy.

Across all functionalities, these findings reveal distinct adoption patterns illuminating counsellors' selective integration of AI assistance based on professional value.
The data suggests a clear hierarchy: interpretive functionalities achieved highest engagement into counselling work, while information extraction tools served primarily orientation purposes.
Accuracy appears fundamental for sustained adoption, with implementation errors undermining otherwise valuable functionalities like the Social Network Graph.
The results suggest that professional counsellors may readily embrace AI tools that stimulate reflection—even when critically evaluating outputs—provided these tools preserve professional autonomy and enhance rather than automate counselling practice.

\section{Conclusion}
This work makes two key contributions to AI-assisted email counselling. First, CAIA demonstrates responsible AI deployment through non-intrusive design principles that preserve counsellor autonomy.
Second, the field evaluation demonstrates that AI functionalities can achieve professional acceptance and usage, indicating which types of AI-generated content may provide meaningful value in counselling practice.

The results demonstrate that interpretive functionalities were primarily adopted and utilised most extensively throughout the evaluation period. 
Information extraction functionalities were utilised, with results indicating their greatest potential value may emerge during case handovers or when extended response times occur between messages.
Qualitative analysis revealed that counsellors valued the non-intrusive design approach, which successfully preserved their professional autonomy. 
However, trust in AI assistance is quickly undermined when errors occur in generated content, highlighting that sophisticated processing pipelines and robust language models are essential for sustained adoption and responsible AI integration into counselling practice.

Several limitations warrant consideration. The study examined professional acceptance rather than counselling effectiveness, the four-week evaluation outside counsellors' habitual platforms may have been insufficient to capture stable adoption patterns and using trained students rather than genuine counsellees may have altered communication dynamics. Additionally, errors in AI-generated content may have distorted adoption patterns.

Future research should address these limitations through controlled effectiveness studies examining counselling outcomes, long-term deployments within established counselling systems, evaluations with genuine counsellees and enhanced technical implementations to minimise accuracy problems.

By examining acceptance patterns, this research demonstrates that AI can offer meaningful value in counselling practice, providing foundations for future research to assess the true effectiveness of AI assistance in improving mental health service delivery.

\section*{Acknowledgements}
The project was funded by the German Federal Ministry of Education, Family Affairs, Senior Citizens, Women and Youth (Bundesministerium für Bildung, Familie, Senioren, Frauen und Jugend) under the funding programme Artificial Intelligence for the Common Good (Richtlinie zur Förderung von Künstlicher Intelligenz für das Gemeinwohl, 2023-2025). 

\bibliographystyle{IEEEtran}
\bibliography{clean_references}

\end{document}